\documentclass[namedreferences]{kluwer}
\usepackage{epsfig,rotate}

\newcommand\kms{km~s$^{-1}$}

\newcommand{\emi}{{\epsilon}}

\newcommand{\fpress}{1}
\newcommand{\fevolve}{2}
\newcommand{\fmodel}{3}
\newcommand{\fcool}{4}
\newcommand{\flpos}{5}
\newcommand{\fprofa}{6}
\newcommand{\fprofb}{7}
\newcommand{\fprofd}{8}

\newcommand{\FIG}[1]{#1}

\newcommand{\PREP}[1]{}

\newcommand{\AnA}[1]{}

\newcommand{\vv}{{\vec v}}
\newcommand{\BB}{{\vec B}}
\newcommand{\JJ}{{\vec J}}

\newcommand{\xx}{{\vec x}}

\newcommand{\ptot}{p_{\mathrm{tot}}}

\begin{document}

\begin{article}
\begin{opening}

\title{Simulations of small-scale explosive events on the Sun}

\runningtitle{Simulations of  explosive events} 

\author{D. E. \surname{INNES}}
\institute{Max-Planck-Institut f\"ur Aeronomie, D-37191 Katlenburg-Lindau,
Germany}
\author{G. \surname{T\'OTH}}
\institute{Dept. of Atomic Physics, E\"otv\"os University, 
           Puskin u. 5-7, 1088 Budapest, Hungary}

\date{\today}

\begin{abstract}
Small-scale explosive events or microflares  occur throughout 
the  chromospheric network of the Sun. They are seen as sudden bursts of
highly Doppler shifted spectral lines of ions formed at  temperatures 
in the range  $2\times10^4 - 5\times10^5$~K. 
They tend to occur near regions of cancelling
photospheric magnetic fields and are thought to be 
directly associated with magnetic field 
reconnection. 
Recent observations have revealed that they have  a bi-directional jet 
structure reminiscent of Petschek reconnection.
In this paper compressible MHD simulations of the evolution of a current sheet
to a steady Petschek, jet-like configuration are computed using the 
Versatile Advection Code. We obtain
velocity  profiles that can be 
compared with recent ultraviolet line profile observations. 
By choosing initial conditions representative of magnetic loops 
in the solar corona and chromosphere, it is possible to explain
 the fact that jets flowing outward into the
corona are more extended and appear before jets flowing towards the
chromosphere. This model can reproduce the high Doppler
shifted components of the line profiles but the brightening at
 low velocities, near the centre
of the bi-directional jet, cannot be explained by this simple MHD model.
\end{abstract}

\end{opening}

\section{Introduction}
The quiet Sun chromospheric network is teeming with tens of thousands of
small-scale explosions. They have a typical lifetime of 60 s, sizes of 
2000 km and  flow velocities of about 100 km s$^{-1}$.
They were discovered in 
high resolution  ultraviolet spectra of the Sun (Brueckner
and Bartoe, 1983; Dere, Bartoe and Brueckner, 1989).
They appear as   small sites with large  red and/or blue Doppler 
shifted emission lines of species which are 
formed at electron temperatures in the range  $5\times10^4 - 5\times10^5$~K.
Recent observations have revealed 
several examples
where  the spatial and temporal evolution of the line profiles
clearly indicate  bi-directional flows (Innes {\it et al}, 
1997; Wilhelm {\it et al.}, 1998).

Several features suggest that the  flows
are quasi-steady and there is continual heat 
input at a local site at the centre of the 
event rather than a real explosion. 
For example,
the position with the brightest emission, which also looks like the central
position of the bi-directional jet, 
moves with a velocity less than 5 \kms\ transverse to the line-of-sight
(Dere, 1994).  
Also
the lifetime of the  events, typically 60 s, is considerably longer than the
cooling time of the   plasma, about 5 s  (assuming  a temperature
$10^5$~K and density $10^{10}$ cm$^{-3}$ (Dere {\it et al.}, 1991)).

Further clues to the flow  come from comparing  line profiles
observed  at the centre of the Sun's disk and towards the limb of the disk.
At disk centre the profiles tend to have  short, bright red-shifted
components and more extended, fainter blue-shifted components.
Also the blue wing tends to emerge first.
The profiles from events occurring  towards the solar limb
have more similar red and blue components (Innes {\it et. al}, 1997).
So it seems as though the extent of the flow and its properties depend
strongly on the conditions surrounding the flow.

There are several suggestions for 
the underlying process triggering these events.
 The first, by Dere {\it et al.}
(1991) and supported by Priest (1998), is that they are 
due to magnetic reconnection between 
network magnetic fields and emerging  loops.
In this case
the observed bi-directional flows would be  reconnection jets.
Another idea is that they are chromospheric evaporation as a result of
coronal microflares (Krucker {\it et al.}, 1997). Yet another idea that 
would also explain the observed pattern of red-blue Doppler shifts,
is that they are
swirling funnels of gas or tornadoes (Pike and Mason, 1998). 
It should be possible to use available high resolution
ultraviolet  spectroscopic observations to judge these suggestions.
To date there have been no theoretical models capable of
predicting emission spectra from any of these scenarios.

In this paper we make a first attempt at computing 
 the behaviour of different temperature   emission
lines from a Petschek-like reconnection scenario (Petschek 1964).
The Petschek solution  is a special  case of fast steady state
reconnection
(Priest and Forbes, 1986; Priest and Lee, 1990)
in which reconnection occurs at a
small  localized X-type neutral point along a narrow current sheet. 
As described by Petschek, material flows into the reconnection point 
and is ejected with
the Alfv\'en velocity in both directions along the neutral sheet.
Although there have been several numerical simulations of this type of 
2-D reconnection ({\it e.g.} Sato and Hyashi, 1979; Biskamp, 1986;
Yan, Lee and Priest, 1991; Jin, Inhester and Innes, 1996)
the conditions in the upper chromosphere/lower transition zone are rather
different to those dealt with in other simulations.
This region is characterized by small,  loop-like structures 
with typical temperatures of a few times $10^4$ K and densities 
about $10^{10}$ cm$^{-3}$ (Dere {\it et al.}, 1991; 
Wilhelm {\it et al.}, 1998). 
Anti-parallel magnetic fields may be driven together 
by, for example, loop footpoint motion, rising loops or by the buffeting effect 
of neighbouring gas motions.

We are not trying to simulate any specific set of observations but to
obtain a realistic feel for the characteristic flow and spectrum 
under different
conditions. 
In section \ref{s-simulation} the simulation models, the boundary and initial
conditions are described. In the next section the basic 
features of the current sheet evolution, the line profiles and 
their temperature dependence are shown.
Finally the strengths and weaknesses of the models are discussed.

\section{Simulation Model \label{s-simulation}}

The evolution is described by the compressible magnetohydrodynamic (MHD)
equations in the following form:
\begin{eqnarray}
{\frac{\partial \rho}{\partial t}}+\nabla\cdot(\vv\rho) & = & 0 
                                                       \nonumber\\
{\frac{\partial \rho \vv}{\partial t}}
+\nabla\cdot(\vv\rho\vv-\BB\BB) +\nabla \ptot & = & 0
               \nonumber  \\
{\frac{\partial E}{\partial t}}
+\nabla\cdot(\vv E+\vv {\ptot} -\BB\BB \cdot \vv) & = & 
                      \nabla\cdot\left(\BB\times\eta\JJ\right)
                                - {Q_{\mathrm{rad}}} \nonumber \\
{\frac{\partial\BB}{\partial t}}+\nabla\cdot(\vv\BB-\BB\vv) & 
 = & -\nabla\times\left(\eta\JJ\right)
\end{eqnarray}
where $\rho$, $\rho\vv$, $E$ and $\BB$ are the density, momentum density,
total energy density and magnetic field,
$\eta$ is the  resistivity coefficient and
$\JJ=\nabla \times \BB$ is the current density.
The magnetic units are chosen so that the magnetic permeability 
(or the $4\pi$ factor of cgs) does not appear in the equations.
The total pressure ${{\ptot}} = p + \BB^{2}/2$. The
thermal pressure $p$ is related
to the total  energy density $E$ by
\begin{equation}
   p=\left(\gamma - 1\right)\left(E-\rho\vv^{2}/2-\BB^{2}/2\right),
\end{equation}
for an ideal gas with adiabatic index $\gamma$.
In these computations $\gamma = 5/3$. 
The temperature units are chosen such that $T=p/\rho$.
The radiative losses are represented by ${Q_{\mathrm{rad}}}$ and the expression 
used is that given in Rosner, Tucker and Viana (1978).
In the solar transition  there are  heat sources, for example thermal
conduction, wave or photon heating, that prevent the
gas cooling below a few times $10^4$ K. These sources are
 ill-defined and  in our simulations we 
simply  decrease the radiative losses to zero at $2\times10^4$~K.
This  means that if the initial temperatures are above $2\times10^4$~K,
the initial state is  not in thermal equilibrium.
In the simulations discussed here the timescale
for cooling from the initial configuration is long compared to the
 simulation timescale.

The MHD equations are solved using a high resolution shock capturing
Total Variation Diminishing Lax-Friedrichs scheme.
This is one of the schemes available in the Versatile Advection Code 
software package\footnote{See http://www.phys.uu.nl/\~{}toth/ for more 
information.} (T\'oth 1996, 1997).
This scheme has been well tested and proved accurate and stable in the
simulation of steady 2-D reconnection (T\'oth, Keppens and Botchev, 1998).
The equations are solved on a non-uniform  2-D Cartesian grid.
Due to the symmetry of the problem, only one quarter of the reconnection
region is represented.
The computational domain $x\in [0,2]$ and $y\in [0,8]$ is resolved by
a $150\times150$ grid. The cell size increases logarithmically 
from $\Delta x=0.0005$ at $x=0$ to $\Delta x=0.1$ at $x=2$. This gives good 
resolution at the neutral point, while the coarser grid at the boundaries 
damps the outward travelling waves without reflection. 

According to the symmetry of the problem,
the boundary conditions are symmetric for $\rho,\ \rho v_x,\ E$, and $B_y$ and
anti-symmetric for $\rho v_y$ and $B_x$ along the $x$-axis, while along the 
$y$-axis the variables $\rho,\ \rho v_y,\ E$, and $B_x$ are symmetric, and
$\rho v_x$ and $B_y$ are anti-symmetric. At the top boundary, all physical 
quantities are extrapolated into the boundary with zero gradient. 
At the right, we fix $\rho,\ \rho v_y$ and $B_y$ to their initial values,
while the inflow mass flux $\rho v_x$ is extrapolated from the outer cell 
on each row so that the inflow speed is determined by the reconnection rate
at the current sheet.
Likewise, the initially zero $B_x$ is extrapolated 
continuously from the last cell row 
so that the field lines can bend in response to the inflow.

The initial configuration of the magnetic field has a current
sheet at $x=0$ with a characteristic width $L$.
The field is parallel to the $y$-axis and varies only along the $x$-axis 
according to
\begin{equation}
  B_y(x) = {\mathrm{tanh}}(x/L).
\end{equation}
In our normalization $B_y=1$ for $x\gg 1$. We also take the density to
be unity at the boundary, thus the Alfv\'en speed $B/\sqrt{\rho}$ is also 
unity.

The initial magneto-hydrostatic equilibrium is maintained by the
total pressure $\ptot$ being constant throughout. Consequently,
the initial distribution of thermal gas pressure $p$ is
\begin{equation}
  p(x) = {\ptot} - B^2_y(x)/2 = p_{\infty} + \frac{1}{2}\mathrm{sech}^2(x/L)
\end{equation}
where $p_{\infty}$ is the thermal pressure at $x\gg L$. The plasma beta at
large $x$ is $\beta=2 p/\BB^2=2 p_{\infty}$.
The initial magnetic and gas pressure across the current sheet are plotted in
Figure~\fpress .

Given the initial magnetic and gas pressure distribution described above, there
is still a free choice of temperature and density distributions.
This allows us to apply this simple current sheet reconnection model
to a range of initial conditions representing different parts of the
solar atmosphere.
For this first study,
the two simplest initial density/temperature configurations are considered: 
a {\em uniform temperature} solution, $T=p_{\infty}$, and 
a {\em uniform density} solution, $\rho=1$. 
Then the density and temperature are given by
$\rho(x)=p(x)/T$ and  $T(x)=p(x)/\rho$, respectively.
The enhancement factor of the density/temperature is
$1+1/\beta$ at the centre of the current sheet ($x=0$)
relative to the value outside ($x>>L$).
Thus along the centre of the current sheet, the density is $1+1/\beta$ 
in the uniform temperature case ($T=\beta/2$) and the temperature is 1/2
 in the   uniform density case ($\rho=1$).

\begin{figure}
\FIG{\epsfig{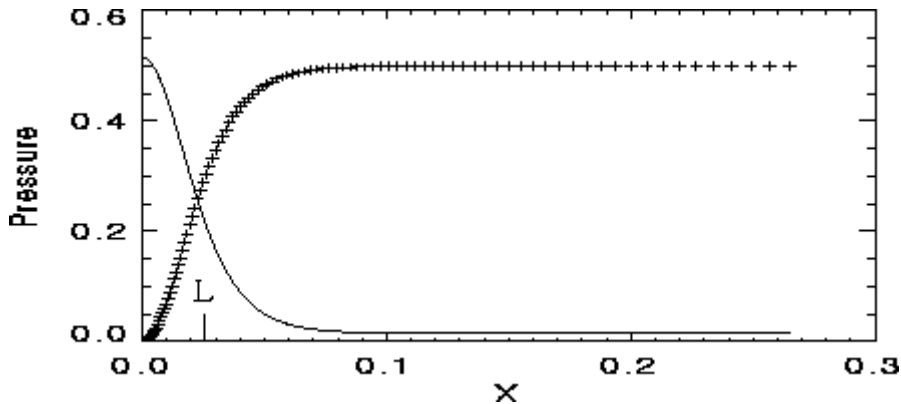}}
\caption{\label{f-press}The initial gas and magnetic (+++) pressure across the
current sheet with characteristic width $L=0.025$. The crosses
are placed at the centre of each  computational cell.}
\end{figure}

In the uniform density case $\beta$ determines the strength of the
shock at the jet head. The jet speed approximately equals
the Alfv\'en speed at large $x$, which is unity, while the
sound speed in the neutral sheet ahead of the jet is 
$\sqrt{\gamma p/\rho}=\sqrt{\gamma(1+\beta)/2}$. 
If $\beta > 2/\gamma - 1=0.2$, the jet is subsonic.
In the final steady solution $\beta$ determines the compression
and opening angle of the jet (Petschek, 1964).
In our simulations we used $\beta =0.03$.

The small central region has an enhanced resistivity that
decreases exponentially from the centre of symmetry by
\begin{equation}
  \eta(\xx)=\eta_0\exp\left[-(x/\ell_x)^2-(y/\ell_y)^2\right],
\end{equation}
where the constants 
$\eta_0=0.0005$ and $\ell_x=\ell_y=0.025$ determine the
resistivity at the origin and the width of the resistivity profile in 
the $x$ and $y$ directions. 

\section{Results}

\subsection{EVOLUTION TO STEADY STATE}
We start from an  equilibrium configuration with two regions
of oppositely directed field lines. 
The localized anomalous resistivity initiates the onset of the reconnection
process between the two regions, and an X-type neutral point forms.
The anomalous resistivity is required to break the initial symmetry
and to maintain the position of the X-point.
In the small region around the X-point, the
current density is high and plasma is accelerated out along the current sheet
(Yan, Yee and Priest, 1992).
In these MHD simulations we do not expect the detailed evolution 
in the vicinity of the X-point to be well modelled. Our aim is to obtain the
basic large-scale flow structure and in   particular the heating, compression
and acceleration behind the slow and fast magnetoacoustic shocks.

\begin{figure} 
\FIG{\epsfig{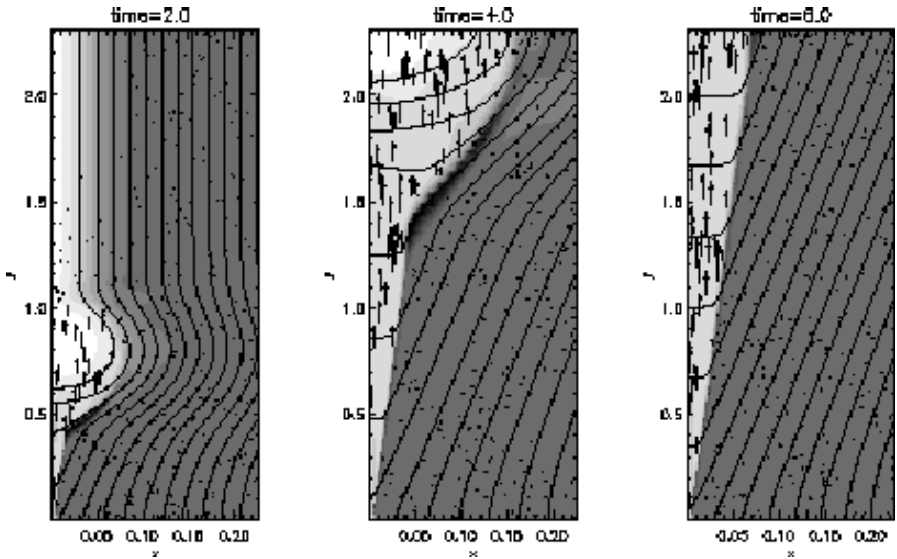}}
\caption{\label{f-evolve}Evolution of uniform density
current sheet. Shown are the temperature (dark tones correspond to low
temperature), velocity vectors, and magnetic field lines.
Only a small part of the full computational domain is plotted
and the $x$ direction is strongly stretched.}
\end{figure}

Outside the diffusion region,
the plasma flows out in narrow jets along the current sheet. 
The evolution for a
low beta, $\beta = 0.03$, and uniform initial density
 is shown in Figure~\fevolve . 
This shows the evolution of the temperature, 
magnetic field and  the velocity vectors after the initial build up of the 
flow around the X-point.
The flow at the jet head 
is supermagnetosonic and there is a fast magnetoacoustic shock.
It shows up as a broad high temperature region with enhanced $B_x$.  
This fast shock was not seen in the compressible MHD 
computations of Sato and Hayashi (1979) 
because they used $\beta = 1$ so the sound speed  along the current sheet
was greater than the Alfv\'en speed outside. 
Once the shock has propagated off the grid
the final system reaches a steady state. 
The gas in the jet  moves at  the Alfv\'en speed along the current sheet.
There is a slow magnetoacoustic or  switch-off 
shock along the boundary of the jet 
that makes an angle $\alpha = 0.025 \approx \eta_o^{1/2}$. 
In the inflowing plasma, the magnetic field decreases from the boundary
at $x \gg 1$ towards the jet while the gas pressure remains almost unchanged.
This is essentially the steady state Petschek solution (Petschek, 1964). 
Other simulations made with different resistivity show that the 
opening angle of the jet depends
on the size of the resistivity and the width of the resistivity function
(Yan, Lee and Priest, 1992).
The structure of the fast magnetoacoustic shock 
depends on the conditions in front of the 
shock.       
Forbes (1986) has studied the structure of stationary 
fast magnetosonic shocks at the
head of reconnection jets directed towards closed magnetic loops tied to the
photosphere. 

\subsection{COMPARISON WITH OBSERVATIONS OF EXPLOSIVE EVENTS}

The chromosphere-corona transition region of the solar atmosphere 
(heights less than 4000 km) is dominated by two temperature regimes.
The gas closest
to the solar surface has a  temperature around $10^4$~K and 
the coronal gas
has a temperature around $10^6$~K. Gas at intermediate temperatures 
cools  rapidly
 and the line emission from ions formed at these temperatures is 
very variable and seems to be coming from small dynamic sites. 
Explosive events are just one well observed example of this variability.

The main evidence that explosive events are bi-directional jets 
comes from a study of their ultraviolet emission line profiles.
Innes {\it et al.} (1997) show examples where  
the line-shifts, corresponding to velocities of
about 100 \kms, change from blue to red within about $2000$ km.
This is what one  would expect if  looking at some small angle
to the flow axis of a bi-directional jet.
Innes {\it et al.} also 
show cases where the length of the region with high line-shifts 
increases during the event lifetime suggesting propagation of
jets into the surrounding material. 
The blue and red  streams often appear to be asymmetric in the sense that,
particularly at disk centre,
the region with blue-shifted emission is more extended and appears 
before the region with red-shifted emission. 

\subsubsection{Structure}

In order to model the explosive events the 
absolute  lengths, density and velocity units are chosen to
be $1000\,$km, $5\times10^{9}\,$cm$^{-3}$ and 
$141\,$km$\,$s$^{-1}$ respectively. Then
the time unit is $7\,$s and  the temperature unit is
$\frac{m}{2 k} v^2_{unit}=1.44\times10^6\,$K,
where $m = 2\times10^{-24}\,$g is the mean mass of atoms,
$k$ is Boltzmann's constant and $v_{unit}$ is the velocity unit.
This gives a temperature about $2\times10^4$~K in the high field region of
the current sheet. The cooling time for  shock heated inflow
gas  is  about 4~s.
The uniform density initial condition has a density of 
$5\times10^{9}\,$cm$^{-3}$
throughout, a temperature of $7\times10^5$ K along the neutral sheet  
and  $2\times 10^4$ K at the inflow boundary.
The other case, the uniform temperature initial condition, has a temperature
of $2\times10^4$ K throughout, a density of $1.7\times10^{11}$ cm$^{-3}$
along the neutral sheet and $5\times10^{9}$ cm$^{-3}$ at the inflow boundary.
These two cases can be thought of as representing magnetic loops in the
corona and magnetic loops in the chromosphere.

\begin{figure}
\FIG{\epsfig{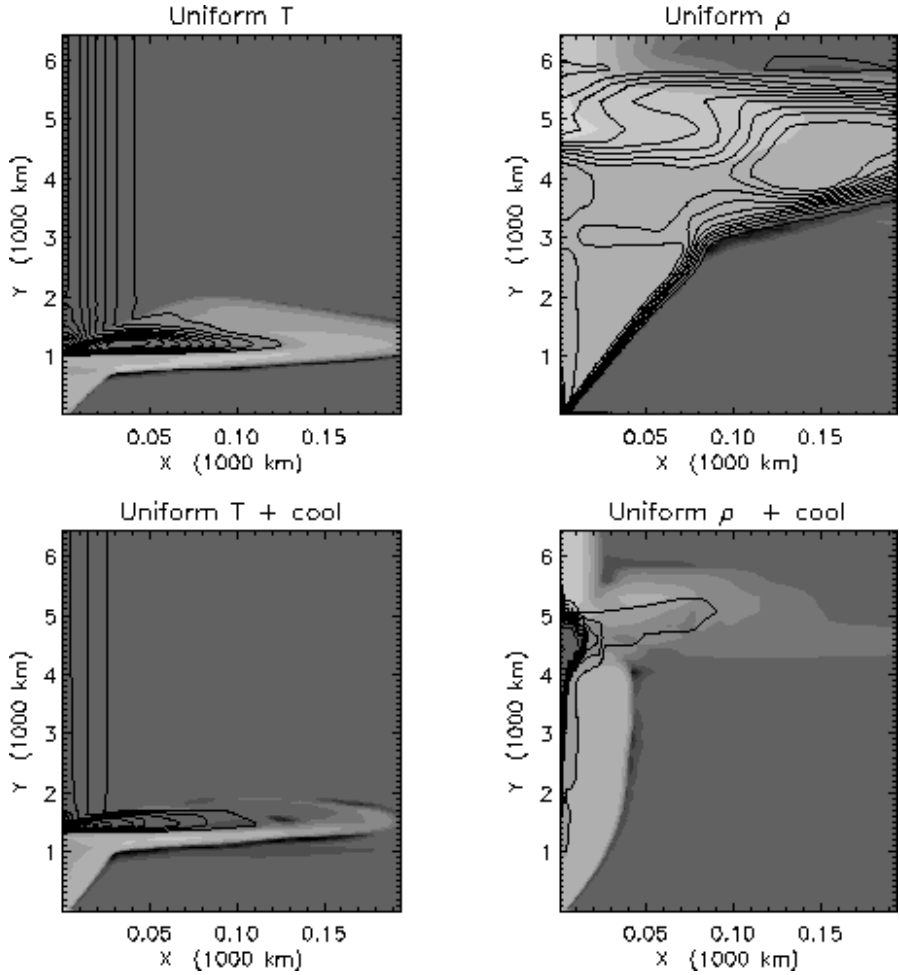}}
\caption{ \label{f-model}
The temperature and density (contours) structure  of the 
reconnection jet at time=50~s for the  uniform temperature and density
initial conditions, without and with radiative cooling.
}
\end{figure}

The temperature and density structure for the two cases with and without
radiative cooling are shown at a time 50~s in Figure~\fmodel .
Here the different evolution scenarios are well illustrated.
The jet moves out at a speed that depends on the square root of the
neutral sheet density.  Thus the jet length is
about 6 times more in the uniform density case.
A natural consequence of this model is that
jets into the hot tenuous corona will be longer and appear before  jets into
the much denser chromosphere.
This may explain why observations made at disk centre
often show more extended blue-shifted than red-shifted emission  and
why the blue-shifted emission often occurs first.
The jet itself on the other hand has the same opening  angle, velocity
and density in both cases.

Radiative cooling considerably changes the structure of the jet.
As shown for the jet from the uniform density initial condition
(right hand panels of Figure~\fmodel), the
jet and its head  are much narrower.
The fast shock at the jet head is confined to the narrow neutral sheet region
and the postshock material cools quickly.
In the case with  uniform temperature  initially 
(left hand panels) the narrowing of the jet is
 not obvious because after 50~s
the jet length is hardly longer than the cooling length.
The temperature and density structure of the cooling jet and their
relationship to the jet flow and magnetic field are shown in
Figure~\fcool .
As in the case without radiative losses, the slow magnetoacoustic
shocks accelerate and heat plasma along the boundary of the jet.
Now the  gas cools as it flows along the jet. The structure along a streamline
is similar to models of 1-D radiative shocks (e.g. Cox, 1972). 
Along the central axis is a cold jet core.
As we shall see, this  shows up as a high velocity, low temperature
component in the line profiles.

\begin{figure} 
\FIG{\epsfig{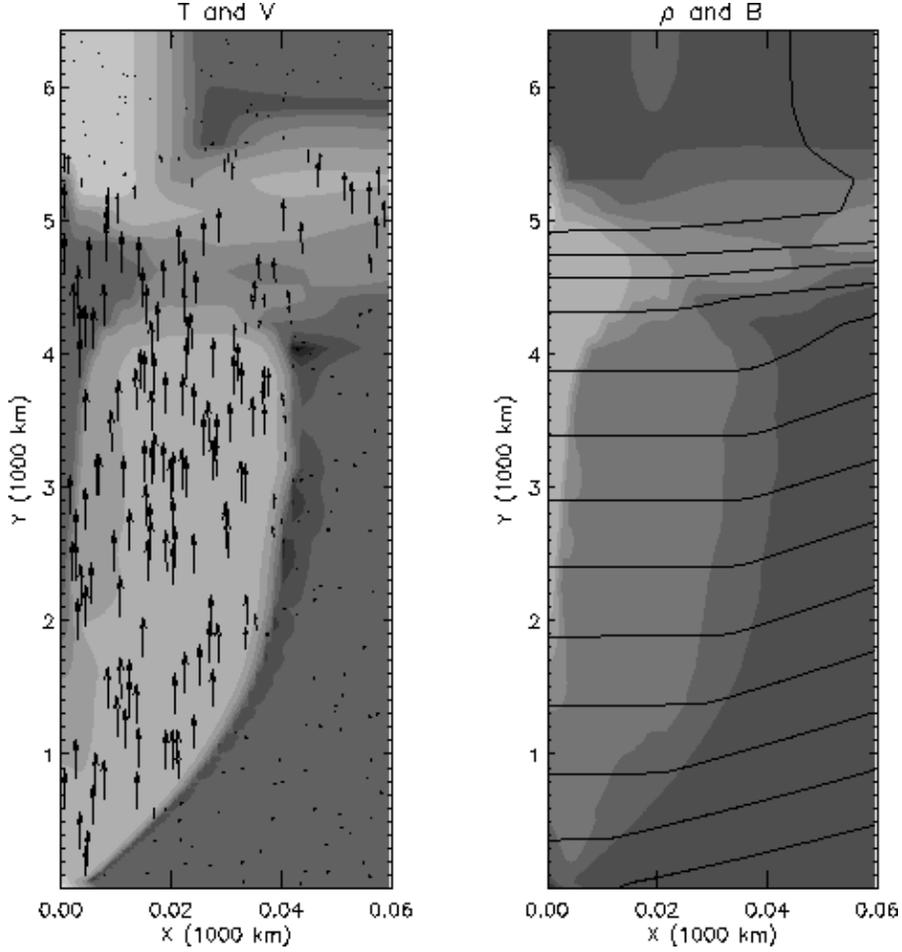}}
\caption{\label{f-cool}The jet structure when radiative cooling is included
in the solution starting from  uniform density. 
On the left  is temperature and velocity and on the right 
density and magnetic field.
}
\end{figure}

\subsubsection{Line profiles}

As we have seen there are three basic regions in the jet structure.
There is the jet source or X-point where magnetic diffusion dominates. Coming
out of the X-point are the reconnection jets. These are
bounded by standing magnetoacoustic shocks and may have cold high velocity
cores.
At the head of the jet is a fast magnetoacoustic shock that moves
outwards from the central source at a speed depending on the 
ram pressure of the jet and the density in the neutral sheet.
These three regions have different spectral characteristics and
under favourable observing conditions it should be possible to 
spatially resolve the regions.

The intensity  of an emission line formed over a restricted temperature range,
$T_{\max}$ to $T_{\min}$, 
is the integral of the emissivity along the line-of-sight,
\begin{equation}
I = \int \int_{T_{\min}}^{T_{\max}} \rho^2 \emi(T) dT dz
\end{equation}
where $dz$ is the line-of-sight path  increment and
the temperature dependence of the line emission is represented
by the function $\emi(T)$.
In our approximation, 
we have assumed that $\emi(T)$ is a simple function defined as
\begin{equation}
  \emi(T) \equiv\left\{\begin{array}{ll}
                    1    & \mbox{if $T_{\min} < T < T_{\max}$}\\
                    0    & \mbox{otherwise.}
                  \end{array}\right. 
\end{equation}
The profiles, shown in Figures~\fprofa --\fprofd , 
are obtained by computing the total
intensity of line emission from gas with a line-of-sight velocity $v$.
Thus the intensity in a given velocity interval $v$ to $v+dv$ is
\begin{equation}
   I_{[v,v+dv]} = \sum_i \rho_i^2 \emi_i dx_i dy_i
\end{equation}
where the summation is over each cell, $i$, in the appropriate region of the
computational domain with line-of-sight velocity
between $v$ and $v+dv$. We used velocity bins of size $dv=1\,$km s$^{-1}$.

The profiles are given for three temperature regimes
$<2\times10^4\,$K,\ $8\times10^4 - 2\times10^5\,$K and
$>5\times 10^5\,$K.
These ranges have been chosen because they outline the explosive event 
 distribution in temperature as inferred from the number of
events seen in various emission lines. 
There are many events seen in lines formed in the middle temperature range, 
a few above $5\times10^5$ K and almost none below $2\times10^4$ K.

As  far as possible we would like to compare the numerical models 
directly to observations therefore we have computed line profiles from 
sections of the models with a scaled length of 2000 km.
This is close to the spatial  resolution of the spectrometer SUMER
(Wilhelm {\it et al.}, 1995, 1997; Lemaire {\it et al.}, 1997), 
the instrument on board SOHO, with which it was possible to spatially        
resolve and observe the evolution of explosive events 
(Innes {\it et al.}, 1997).
Solutions with different temperature/density initial conditions differ
only in the fast shock structure at the jet head and the
propagation speed of the jet but not in the jet structure itself.
Therefore we use the solution from the uniform density
initial condition to show the evolution of the jet. 
In this simulation the jet moves a distance
of 5000 km in 50~s. The different regions of such a jet 
fall conveniently into   2000 km spatial sections.
For the other initial condition, 
only the profiles from the full jet structure are shown. 

The profiles from  the uniform density evolution are displayed in Figures
\fprofa\ and \fprofb . These figures are organized so
that each row corresponds to a different temperature, each column 
to a different section of the grid and different plot line styles
distinguish profiles at different times.
The spatial sections are marked on Figure~\flpos\
and labelled `jet source', `jet' and `jet head'.
These labels  describe the structure at 50~s. At earlier
(or later) times different regions of the jet fill these sections
of the computational grid.
For example at 30~s the jet
 head is in the `jet' section.
The `jet source' region covers both the up- and downward flowing jet.
For this region the profiles are computed after reflecting the structure
from the lower 1000 km across the $x$-axis. This makes it easier to see the 
relative intensities in the high and zero velocity components at the
centre of the bi-directional jet.
All profiles are shown for a viewing angle of $45 $ degrees.
Thus a jet flowing along the axis at the Alfv\'en speed at $x>>L$
has a line-of-sight
velocity 100 km s$^{-1}$.

\begin{figure}
\FIG{\epsfig{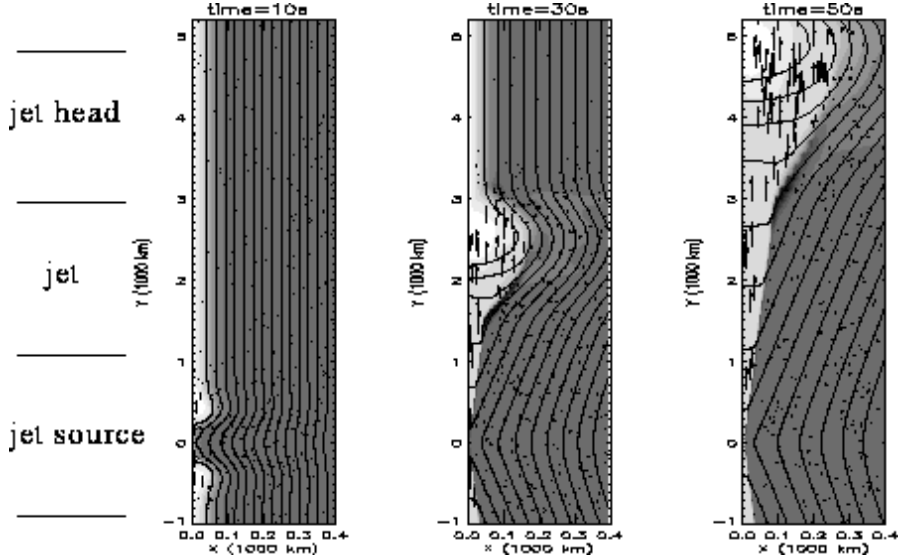}}
\caption{ \label{f-lpos}
The uniform temperature jet structure at the three times where line
profiles are calculated. The spatial sections over which the profiles 
have been integrated are marked. The names are  relevant for
the  structure at time 50~s.}
\end{figure}

High velocity emission is seen in the middle and high temperature range for 
the evolution without radiative cooling.
At middle temperatures both the jet and the head emit.
In this simulation and at this viewing angle (45 degrees), the profiles
appear broad.
The velocity  of the emission from the jet head is smaller than
from the jet because the head emission is coming from the sides
of the jet head where the flow velocity is less than the on-axis flow. 
The shock at the
jet head is about a factor $2$ brighter than the jet.
This ratio depends on the strength of the
fast shock at the jet head.
For higher values of $\beta$ the shock and  the jet emission are weaker.
The high temperature emission 
is coming from the central front of the fast shock and
only the jet head, not the jet itself shows up in hot lines.
At $T<2\times10^4$ K there
is no high velocity plasma. The emission is coming from a thin region just
upstream of the slow mode shock.

\begin{figure}
\FIG{\epsfig{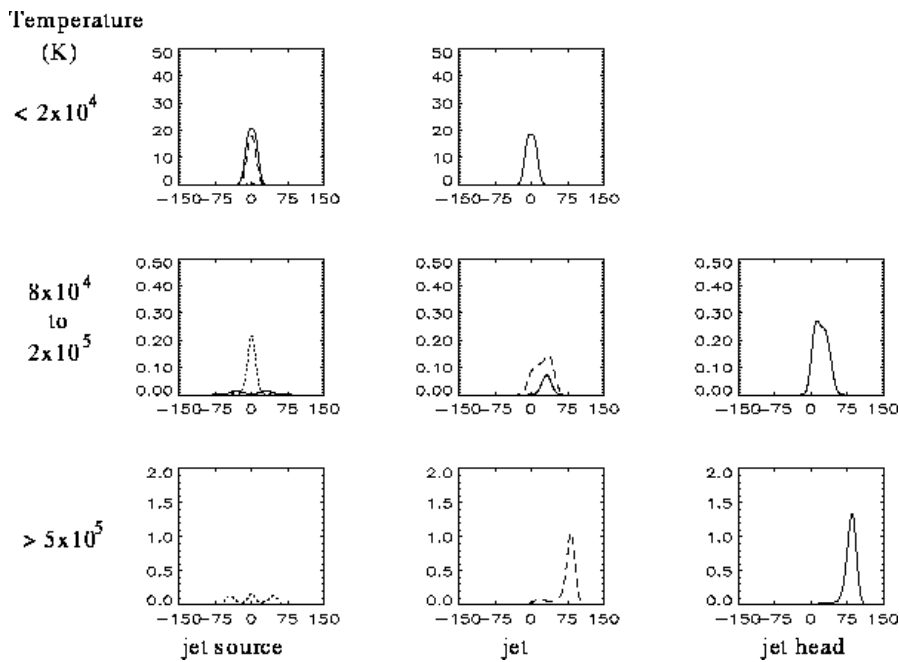}}
\caption{ \label{f-profa}
Line profiles (relative intensity vs. velocity in km s$^{-1}$)
from the uniform density initial condition. Each row shows line profiles at the
temperatures given on the left, from the
three sections labelled on the bottom and marked in Figure~\flpos .
The different line profiles correspond to different times.
t = 10~s (dotted), t = 30~s (dashed) and t = 50~s (solid).
The line-of-sight is 45 degrees to the jet axis.
}
\end{figure}
\begin{figure} 
\FIG{\epsfig{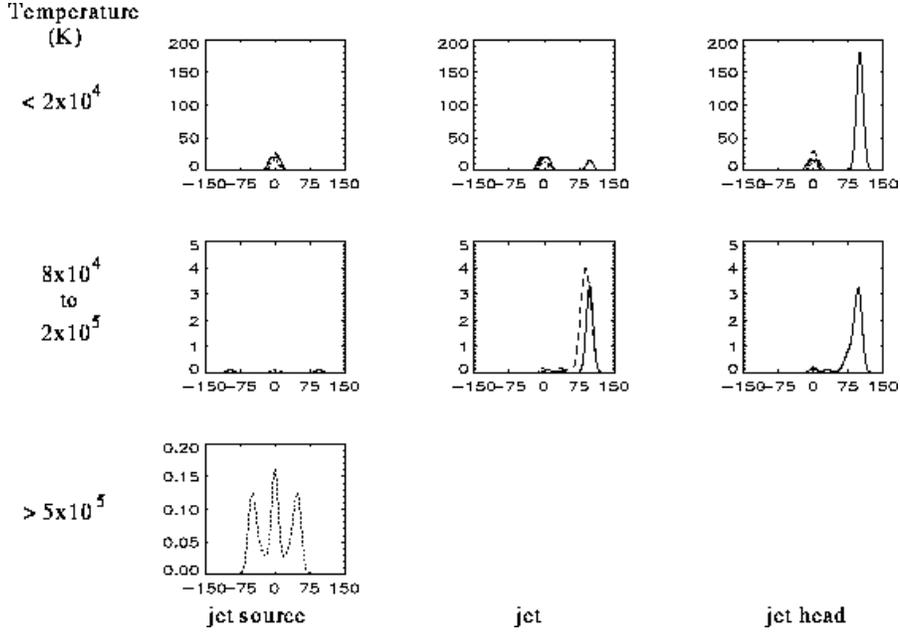}}
\caption{\label{f-profb}
Line profiles from the uniform density condition with cooling.
The lay-out of the figure is as described in Figure~\fprofa .
}
\end{figure}

As shown in Figure~\fprofb ,
when radiative cooling is included in the computations, the high temperature
emission disappears and cold high velocity profiles are
obtained due to the formation
of the cold jet core.
As the head moves out the intensity of the cold temperature 
line increases ({\it c.f.} right panel of the top row).
In the middle temperature range high 
velocity emission is produced along the jet and near the centre of the 
fast shock. 
In the example illustrated here, there is also a weak 
lower velocity component coming from
flows along the outer edge of the jet head. 
As in the simulation without radiative cooling, the relative 
strength of the emission from the jet head and jet 
depends on the flow structure at the jet head.
This in turn depends strongly on initial conditions.

\begin{figure}
\FIG{\epsfig{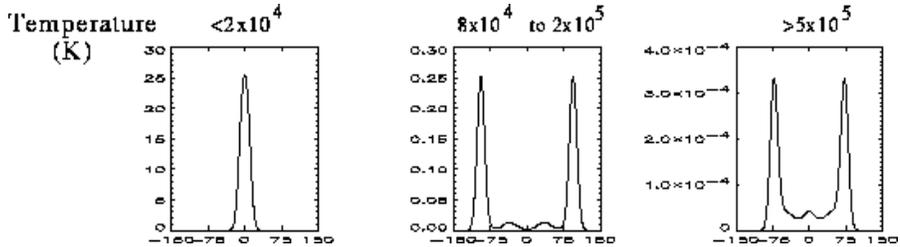}}
\caption{ \label{f-profd}
Line profiles at three different temperatures at $t = 50\,$s,
from the current sheet starting with a uniform temperature.
This shows the total jet profiles from the structure reflected across the
$x$-axis.
}
\end{figure}

The final set of profiles, given in Figure~\fprofd , are from the
uniform temperature initial condition with radiative losses at $t = 50\,$s.
At this time, as can be seen in Figure~\fmodel , the jet head has only 
moved
1 length unit (1000 km) so the profiles shown here are the profiles
from the full structure after it has been reflected across the $x$-axis.
The profiles from the equivalent structure in the evolution of the
uniform density current sheet are the dotted lines ($t = 10\,$s) in 
Figures \fprofa\ and \fprofb .

The low temperature profiles are dominated by  emission from the
plasma in the high magnetic field region of the current
sheet.
The jet emission produces only very low
intensity wings on the original zero velocity
profile. At middle temperatures the high velocity components are
strong and have a double component structure. Again the highest velocities
are coming from the jet and the lower velocity component from the jet head.
The high temperature brightening is very weak. The emission is mostly coming
from the gas on-axis just behind the fast shock. The velocity is slightly
less than the jet velocity.

\section{Discussion}

Although these models are gross simplifications of
the solar  structure they illustrate a few basic
features of the emission from reconnection jets. 
(1) High velocity components in lines formed at temperatures
$\approx~10^5$ K, are emitted from the jet irrespective
of initial conditions.
(2) The jet length depends on the ram pressure (which is controlled by the
reconnection rate) and the density in front of the jet.
(3) There is no brightening in  zero velocity emission either along the
jet or at the reconnection point.
(4) Due to radiative cooling the jets have cold cores and this may
produce high Doppler shifts in cold lines.
(5) Multiple components or broad line profiles 
may be  obtained in lines  formed $\approx~10^5$ K
due to different flow velocities at the jet head and
along the jet.
Features such as the ratio of the line intensity in the jet head and jet
are very dependent on initial conditions and cannot be used as a test for the 
model.
Features (1), (2) and (5) are consistent with observations but (3) and
(4) are not.

High velocity emission in cold lines, feature (4), may actually
be present in the data but weak compared to the background because
the filling factor of the jets is very small. In a $10^9$ km$^3$
volume the jet would occupy at most $10^{-4}$ of the space. 
By careful analysis of SUMER data,
it should be possible to obtain
upper limits to the amount of cold high velocity material and thus obtain
a better assessment of the model.

The other failing is maybe more serious but even so there is no 
definitive reason
to reject the basic model. The diffusion region is not well 
represented by the 2-D MHD simulations.
In  3-D simulations the diffusion region
breaks up into many small high density islands (Otto, 1998). 
These would most likely
generate bright broad profiles in lines formed around $10^5$ K.
Also particle simulations have shown that electrons are easily 
 trapped in the diffusion region where they are rapidly accelerated
(B\"uchner and Kuska, 1997).
These high energy electrons may heat the surrounding low velocity
inflowing plasma.

The main success of the model is that due to the presence of 
slow magnetoacoustic shocks along the jets' boundary, it is able to explain
the almost stationary red and blue shifted emission at high 
(supersonic) velocities in lines with formation temperatures 
around $10^5$~K for a period much longer than the cooling time of the gas.

\acknowledgements
We thank the Max-Planck-Institut f\"ur Aeronomie and the Astronomical
Institute at Utrecht for financial support during our mutual visits. 
G.T. developed the Versatile Advection Code as part of the project on
`Parallel Computational Magneto-Fluid Dynamics', funded by the
Dutch Science Foundation (NWO).
G.T. is currently supported by a Hungarian Science Foundation (OTKA) 
postdoctoral fellowship (D~25519) and receives partial support from the OTKA
grant F 017313.

\end{article}

\end{document}